\documentstyle[aps,preprint]{revtex}
\begin{document}
\preprint{KSUCNR-004-96, INT96-00127}
\draft
\title{The hadron-quark transition with a 
lattice of nonlocal confining solitons}
\author{Charles W. Johnson\thanks{
electronic mail: johnson@ksuvxd.kent.edu}
and George Fai\thanks{
electronic mail: fai@ksuvxd.kent.edu}
}
\address{Center for Nuclear Research, Department of
Physics, Kent State University, Kent, OH 44242}
\author{Michael R. Frank\thanks{
electronic mail: frank@phys.washington.edu}
}
\address{Institute for Nuclear Theory, University of Washington,
Seattle, WA 98195}

\maketitle
\begin{abstract}
We use a lattice of nonlocal confining solitons to describe nuclear
matter in the Wigner-Seitz approximation. The average density is varied 
by changing the size of the Wigner-Seitz cell. At sufficiently large
density quark energy bands develop. The intersection of the filled valence 
band with the next empty band at a few times standard nuclear density
signals a transition from a color insulator to a color conductor
and is identified with the critical density for quark deconfinement.
\end{abstract}
\pacs{}

The search for the hadron-quark transition in nuclear collisions has
provided powerful motivation for experimental and theoretical research
in high-energy nuclear physics. Experimental results are now emerging
from Brookhaven's Alternating Gradient Synchrotron (AGS) and CERN's 
Superconducting Proton Synchrotron (SPS), while the planned Relativistic
Heavy Ion Collider (RHIC) at Brookhaven represents the
first major construction priority of the US 
nuclear physics community and is expected to be operational before the
turn of the century. Work on the theory side can be broadly categorized 
into studying the fundamental properties of the strong interaction 
and exploring 
phenomenological aspects such as the signatures and characterization
of the quark-gluon plasma (QGP). While phenomenological support for experiment
is crucial for the success of the overall effort, in the present paper 
we address the more fundamental question of modeling 
the hadron-quark transition.

Modeling the expected transition will remain important as long as a 
complete solution of quantum chromodynamics (QCD) in a reasonably large 
space-time volume is out of reach. Soliton matter was proposed as a model of 
dense nuclear matter by several 
authors\cite{banerjee85,reinhardt85,achtzehnter85}.
In the following we focus attention on the behavior of soliton matter
as a function of density at zero temperature.
The simplest approach in this context is to think about nuclear matter 
as a crystal of MIT bags\cite{zhang86}. In more elaborate many-soliton 
models the rigid walls of the MIT bags 
are replaced by external fields which are supposed to represent the collective 
nonperturbative effects of QCD.\cite{birse88} 
Several aspects of this picture were 
developed further for the case of a single soliton in the framework
of the Color Dielectric Model (CDM)\cite{fpw88}. 

The alternative approach taken here is based on the 
hadronization\cite{cahill92,frank95b} of a model truncation of QCD, 
known as the Global 
Color Model (GCM)\cite{cahill85}.  The GCM is defined by the action
\begin{equation}
S_{GCM}[ \overline{q},q]  = \int d^{4} x d^{4} y
\biggr\{ \overline q(x)\biggr[ (\gamma \cdot \partial +  m_{q})
\delta(x-y) \biggr] q(y) 
+ \frac{g^2}{2} j^a_{\nu}(x)D(x-y)j^a_{\nu}(y) \biggr\} \, ,\label{gcm}
\end{equation}
where $j^a_{\nu}(x) = \overline{q}(x)\gamma_{\nu}\frac{\lambda^{a}}{2}q(x)$ is
the color quark current.  QCD phenomenology such as confinement is 
incorporated into the quark-quark interaction, $D$.\footnote{For convenience 
a gauge is chosen here for which the effective gluon 
two-point function is diagonal in Lorentz indices,
$D_{\mu\nu}(x-y) = \delta_{\mu\nu}D(x-y)$, and the Euclidean metric 
is employed.}
This model has proven successful 
in the description of low-energy QCD through the reproduction of 
chiral perturbation theory\cite{cahill85,xpt}, 
meson form factors\cite{formfactors}
and spectra\cite{frank95a,frank96a}, and the soliton\cite{gsm} and 
Fadeev\cite{cahill92} descriptions of the nucleon.  The relationship 
of this model to other models of QCD has recently been 
summarized\cite{cahill96}.

The formation of solitons is achieved by minimizing the energy functional 
for the system at the mean field level 
with the baryon number constrained by a chemical potential.  For simplicity, 
as with the previous treatment of this model\cite{gsm}, we restrict our 
attention here 
to a scalar-isoscalar mean field.  The self-consistent solution of the 
mean-field equations entails iterating to convergence the Dirac equation for 
the vacuum-dressed quarks and a nonlinear Klein-Gordon 
equation for the scalar field.

There are two novel features of this model that are particularly relevant 
to the present discussion.  First, the meson fields and their self 
interactions are generated through the hadronization of Eq. (\ref{gcm}).  
This implies in particular that the ``macroscopic" behavior is determined 
entirely from the underlying quark-quark interaction and 
parameters which enter at the quark level.  Second, the confinement 
mechanism that arises here is due to vacuum repulsion as supplied by the quark 
self-energy dressing.  This point is demonstrated as follows. 

The quark-quark interaction, $D$ in Eq. (\ref{gcm}), can be chosen to produce 
a quark self energy, $\Sigma (p)=i \gamma \cdot p [A(p^2)-1]+B(p^2)$, which 
prohibits the propagation of quarks in the vacuum.  Here we take for 
example in momentum space $g^2D(q)=3\pi ^4\alpha ^2\delta^{(4)}(q)$, which 
in the rainbow approximation\cite{review} to the Dyson-Schwinger equation 
gives the result
\begin{eqnarray}
A(p^2) = \left\{ \begin{array}{c} 2 \\
	 \frac{1}{2}[ 1 + (1+\frac{2\alpha^2}{p^2})^{\frac{1}{2}}]
		     \end{array}  \right. \;\; , \; & 
B(p^2) = \left\{ \begin{array}{c} (\alpha^2-4p^2)^{\frac{1}{2}} \\
	                               0
		     \end{array}  \right. \;\; & 
		 \begin{array}{c} p^2 \leq \frac{\alpha^2}{4} \\
				  p^2   >  \frac{\alpha^2}{4}	 		
		     \end{array}	\;\;\; .    \label{AB}
\end{eqnarray}
That the forms (\ref{AB}) produce a model of confinement as described 
above can be seen by the absence of a solution to the equation
$p^2+M^2(p^2)=0$, with $M=B/A$ (no on-mass-shell point).  

The mean field produces a 
cavity in the vacuum in which the quarks can propagate.  For example, 
in the case of quarks coupled to a {\it constant} scalar mean field by the 
scalar self-energy function\footnote{This coupling actually arises from 
chiral symmetry considerations\cite{cahill85}.} $B$, the quark 
inverse Green's function obtains the following form: \newline
$G^{-1}(p)=i \! \not \! pA(p^2)+B(p^2)(1+\chi )$, where $0\geq \chi 
\geq -2$ characterizes the strength of the mean field.  In this case a 
{\it continuous} single-particle energy spectrum, $E^2({\bf p}^2)=
{\bf p}^2+M_C^2$, is obtained because of the constant potential, 
with the constituent mass given by 
\begin{equation}
M_C^2=\frac{\alpha ^2}{4}\frac{(1+\chi )^2}{1-(1+\chi )^2} \,\,\, .\label{M}
\end{equation}
From (\ref{M}) it is evident that the increase of the constituent mass 
as the strength, $\chi $, of the mean field decreases toward its vacuum 
value ($\chi =0$), is due to the repulsive interaction with the vacuum.  
For the case of the constant mean field, the quarks are allowed to propagate 
throughout space.  However, due to the energy stored in the mean field, 
the self-consistent (minimum-energy) solution acquires a finite range. In 
that case, as a quark in the system is separated from the others, the 
influence of the mean field on the quark diminishes and the mass rises as 
in (\ref{M}).  The quarks are thus confined to the region of nonzero mean 
field by virtue of their interaction with the vacuum.  

In the description of matter, as the density increases the space between 
isolated solitons decreases, and the mean field can acquire nonzero values 
throughout space.  One can conclude based on Eq.(\ref{M}) that as the 
vacuum is filled by the nonzero mean field the quarks can sample a larger 
region of space.  Finally, in sufficiently dense matter, 
localized clusters of quarks (nucleons) cease to 
be energetically favored, and a conducting phase is established.  

To discuss the many-soliton problem in more detail
and to identify the transition density, we now consider a 
lattice of static solitons as described above.  
The relative motion of the nucleons is thus neglected in the
present treatment and in similar studies. Here we are not interested in
reproducing the saturation properties of nuclear matter. Rather, we
focus on qualitative changes in the behavior of the system (as a function of
the lattice spacing) that can be identified with the hadron-quark transition 
in the framework of the model. A more complete treatment should include
the kinetic energy of moving solitons. 

As opposed to the single-soliton situation, where the boundary conditions
require that the self-consistent solutions of the Dirac and Klein-Gordon
equations vanish as $r \longrightarrow \infty$, we have periodic
boundary conditions on the lattice,
$\sigma({\bf r}) = \sigma({\bf r} + {\bf a})$   
for the meson field,
and similarly for the quark spinors, where $\bf{a}$ is the lattice vector.
As is well known from solid state physics\cite{kittel}, for sufficiently small 
values of the lattice constant $a = |{\bf a}|$,
the sharp energy eigenvalues of the isolated soliton
are replaced by sets of closely spaced eigenvalues (energy bands),
each soliton contributing one energy level to each band. 
As the lattice constant goes to infinity, the bands
converge to the single-soliton solutions.

Following earlier calculations for soliton matter\cite{banerjee85,birse88},
we address the problem in the Wigner-Seitz approximation\cite{wigner33}. 
Introducing a spherical Wigner-Seitz cell of radius $R$ permits the 
decomposition of the quark spinors as
\begin{equation}
u_{n\kappa}(r) = \left( \begin{array}{c} 
			g_{n\kappa}(r){\cal Y}_{J\ell}^{m_J}({\hat r}) \\
			f_{n\kappa}(r){\cal Y}_{J \bar{\ell}}^{m_J}({\hat r}) 
			\end{array} \right)	\;\;   ,
\end{equation}
where the $\cal Y$-s are vector spherical harmonics, and prescribes the
following boundary conditions at the edge of the Wigner-Seitz cell:
\begin{equation}
\left. \frac{d}{dr}\sigma(r)\right|_{r=R} = 0
\end{equation}
for the sigma field, and 
\begin{equation}
\left.\left. \frac{d}{dr}g_{n\kappa}(r)\right|_{r=R} = 
f_{n\kappa}(r)\right|_{r=R} = 0              \label{qbound}
\end{equation}
for the upper and lower components of the quark spinors corresponding to 
the lowest-energy state in each energy band\cite{banerjee85,birse88}. 

To solve the coupled equations for the many-soliton problem, it is 
advantageous to work with the Fourier expansions of the radial quark wave 
functions $g_{n\kappa}(r)$ and $f_{n\kappa}(r)$. The Dirac equation then 
reduces to a matrix equation in momentum space and the boundary conditions
can be easily incorporated. We start by solving for the quark energy 
eigenvalues in an arbitrary sigma field. The quark wave functions of
the lowest-energy state are then generated, which (after Fourier
transformation) provide the source 
of the sigma field as input to the nonlinear Klein-Gordon equation in 
coordinate space. The Dirac equation is solved again in the new sigma field
and the quark wave functions are compared (in quadrature)
to the ones obtained in the previous iteration. Dependent on the 
radius of the Wigner-Seitz cell, $R$, and on the required tolerance,
the procedure typically converges in 3-6 iterations. Once the final
sigma field has been obtained at a given $R$, the Dirac equation in this field
is solved for the energies of the higher excited states. 
Each Dirac spinor carries two different
orbital quantum numbers ($\ell$ and $\ell'$) and only the total angular
momentum $j$ is a good quantum number. In particular, the lowest-energy state
has $\ell=0, \; \ell'=1, \; j=1/2$, and the next state is characterized by 
$\ell=1, \; \ell'=2, \; j=3/2$.\cite{bhaduri88} We will refer to these states 
as the $1s1/2$ and $1p3/2$ states, respectively.

The bottom and top energies of the lowest energy bands as a function of 
the radius of the Wigner-Seitz cell for the parameter value $\alpha=1.35$ GeV 
in (\ref{AB}) are shown in the Figure. 
The different symbols
represent the calculated energies of the bottom of the three lowest 
energy bands. On one point we indicate a typical uncertainty we associate with 
our calculation. This uncertainty mainly arises from the freedom in 
prescribed tolerances at different stages of the calculation.
The calculated points are fitted to smooth curves to guide the eye 
and to approximately determine the intersection point between the top
of the lowest energy band ($1s1/2$) 
and the bottom of the next two bands ($1p3/2$ and $2s1/2$).
For the top of the energy bands $\epsilon_{top}$ as a function of the cell 
radius $R$ we use the simple approximation\cite{banerjee85}
\begin{equation}
\epsilon_{top} = (\epsilon_{bot}^2+(\frac{\pi}{2R})^2)^{1/2}   \;\; ,
\label{top}
\end{equation}
where $\epsilon_{bot}$ is the bottom of the energy band obtained with 
the boundary conditions (\ref{qbound}).
More elaborate prescriptions for the top of the bands are also used in the 
literature\cite{birse88}, but the above simple estimate is consistent 
with the qualitative nature of the present study. For large $R$, 
$\epsilon_{top} \longrightarrow \epsilon_{bot}$, and the eigenvalues 
approach those of the isolated soliton. As $R$ is decreased, we first 
observe a decrease in the energy of the lowest band. This attraction
is a consequence of the boundary condition on the large component of 
the quark wave function (\ref{qbound}), which allows less curvature and
therefore less quark kinetic energy than in the case of the isolated
soliton. For higher densities, when the solitons and the quark wave 
functions start to significantly overlap, the resulting repulsion 
overcomes the attraction, and a minimum develops. Since we have a
mean-field model lacking the details of the nucleon-nucleon interaction,
we do not expect the minimum to be at the saturation density of
nuclear matter. Upon further increase of the density, 
the energy of the $1p3/2$ band does not increase any longer; this is
attributable to the increase of the sigma field at the cell boundary,
which makes it advantageous for the system to arrange itself in a 
configuration with nonzero orbital angular momentum and with quark 
density peaked away from the origin.

The intersection of different bands can be interpreted to
signal a transition to a qualitatively different high-energy phase. 
For moderate values of $R$ no bands intersect, and matter is an insulator.
However, as $R$ decreases, the lowest, occupied band intersects 
the $1p3/2$ and $2s1/2$ bands. When the energy 
of an unfilled level falls below that of the highest occupied state,
quarks with this energy become free to migrate throughout the crystal as 
electrons in a metal, and color conductivity sets in. We follow Ref. 
\cite{banerjee85} and identify the onset of color conductivity with the
deconfinement transition. The value of the critical density $\rho_c$ in the
model depends on the assumptions on band filling.\cite{birse88}
Taking the simple approximation of uniformly filled bands and
considering the $1s1/2$ --- $1p3/2$ band crossing, with 
$\alpha=1.35$ GeV we get $\rho_c = (2.6 \pm 0.2) \rho_0$,
where $\rho_0 = 0.17$ fm$^{-1}$ is the standard nuclear matter density. 
Assuming that the lowest energy band is partially filled will increase
the critical density obtained in the model.

In the Table we show variations corresponding to changes in the value of the 
parameter $\alpha$ in Eq. (\ref{AB}). The root mean square radius of the 
soliton for large $R$ (isolated nucleon), the Wigner-Seitz radius 
and density belonging to the minimum energy, the critical Wigner-Seitz radius,
and the critical density are displayed for selected values of $\alpha$. 
The present model does not include any meson dressing, and thus care
should be taken when the calculated $\langle r^2 \rangle^{1/2}$ is
compared to the experimental value of the root mean square charge 
radius of the proton, $\langle r^2 \rangle^{1/2}_{exp} \approx 0.83$ fm.
We feel that in this schematic model the range selected represents a
reasonable interval of variation for the single free parameter $\alpha$.

In summary, we discussed the properties of an infinite system of 
nonlocal, confining solitons in the Wigner-Seitz approximation. Due to the 
internally-generated meson fields and the specific confinement mechanism 
associated with the model, our study represents a step forward in the
description of nuclear matter as a system of solitons. The quark
energy bands that develop in the system with decreasing cell size were
calculated. We identified the insulator-conductor transition in the 
crystal with the deconfinement transition expected in dense hadronic matter.
In studies of the 
nuclear equation of state it is customary to use a hadronic equation of state
at low energy densities and a bag-model equation of state at high 
energy densities; the critical parameters are than determined by extrapolating 
both descriptions to the phase transition region and solving the Gibbs 
conditions for phase coexistence\cite{mfco93}. The present model is 
much more satisfactory as it uses the same basic
degrees of freedom in both phases and across the phase transition.
In the future, the model can also be used to examine in-medium properties
such as the dependence of the nucleon mass on the density of the medium.

\acknowledgements

Two of us (C.W.J. and G.F.) are grateful for the partial support and 
hospitality of the Institute for Nuclear Theory at the University of 
Washington. This work was supported in part by the Department of
Energy under Grant Nos.\ DOE/DE-FG02-86ER-40251 and  DE-FG06-90ER-40561.

\newpage
\begin{center}
{\bf Figure caption}
\end{center}

The bottom and top energies of the lowest energy bands of the soliton 
lattice as a function of the radius of the Wigner-Seitz cell for
$\alpha=$1.35 GeV. The symbols represent the calculated energies of the
bottom of the three lowest energy bands. For the top of the energy bands
the approximation %(\ref{top}) 
(7) is used.

\begin{center}
{\bf Table caption}
\end{center}

Variations of the root mean square radius of a single soliton, the 
Wigner-Seitz radius and density belonging to the minimum energy, 
the critical Wigner-Seitz radius, and the critical density with the single
model parameter $\alpha$.

\newpage
\begin{center}
%\begin{minipage}{7 in}
\begin{tabular}{||l||c|c|c||}  \hline \hline
$\alpha$ (GeV) 			  & 1.25 & 1.35 & 1.45 \\  \hline
$\langle r^2 \rangle^{1/2}$ (fm)  & .71  &  .67 &  .64 \\
$R_{min}$ (fm)          & 1.33$\pm$ .02 & 1.26$\pm$ .02 & 1.17$\pm$ .02   \\
$\rho_{min}/\rho_0$     & .60 $\pm$ .03 & .70 $\pm$ .04 & .88 $\pm$ .05   \\
$R_c$ (fm)		& .86 $\pm$ .02 & .82 $\pm$ .02 & .77 $\pm$ .02   \\ 
$\rho_c/\rho_0$		& 2.2 $\pm$ .2  & 2.6 $\pm$ .2  & 3.1 $\pm$ .3    \\ 
\hline \hline
\end{tabular}
%\end{minipage}
\end{center}

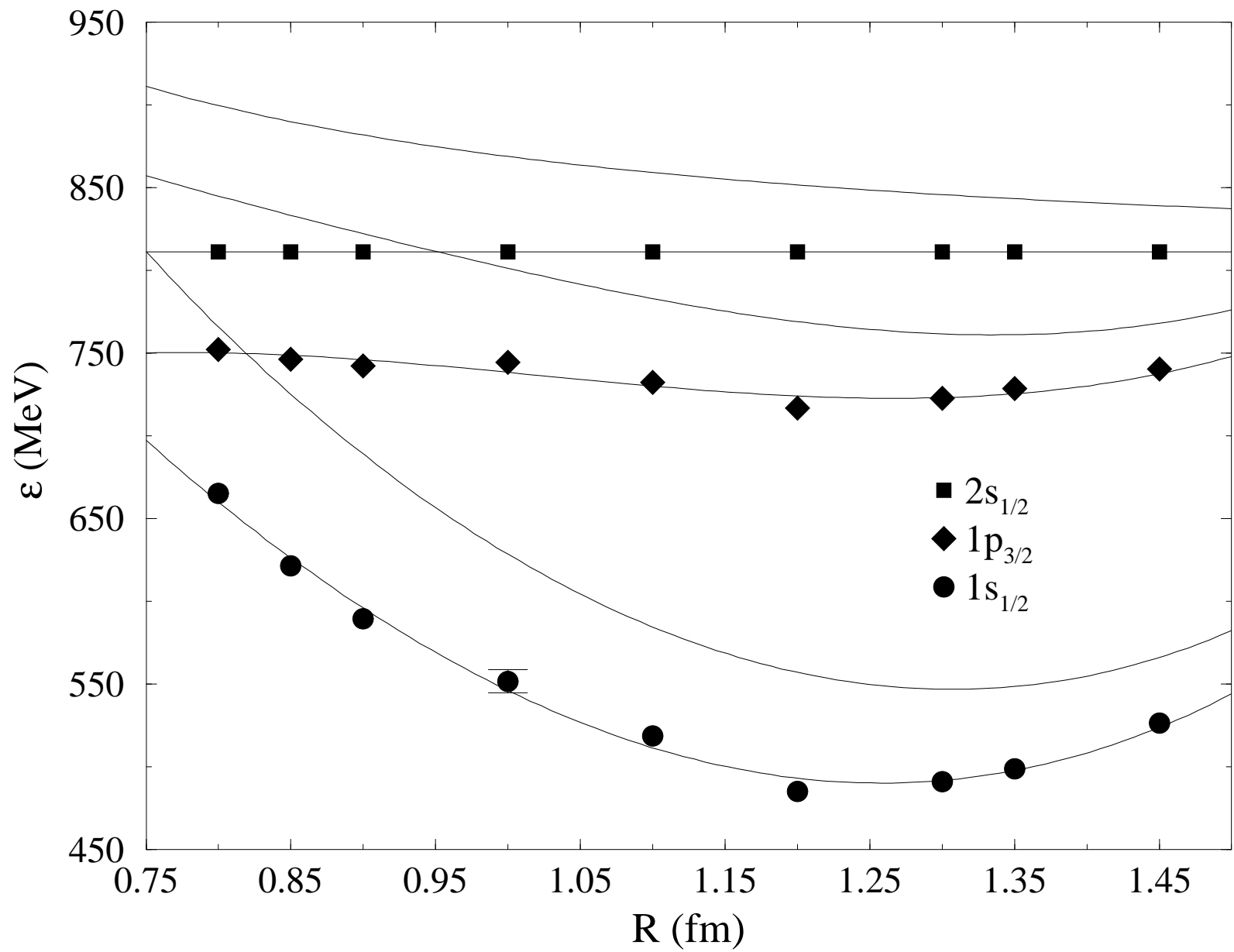
\end{document}